\documentclass[a4paper]{jpconf}
\usepackage{graphicx}
\usepackage{amsmath,amssymb}
\usepackage{ascmac}
\usepackage{subfigure}
\newcommand{\dr}{\ms\hline \ms \hline\ms}
\begin{document}
\title{Millikelvin LEED apparatus: a feasibility study}

\author{K. Matsui, S. Nakamura, T. Matsui, and Hiroshi Fukuyama}

\address{Department of Physics, Graduate School of Science, The University of Tokyo,
7-3-1 Hongo, Bunkyo-ku, Tokyo 113-0033, Japan}

\ead{kmatsui@kelvin.phys.s.u-tokyo.ac.jp, hiroshi@phys.s.u-tokyo.ac.jp}

\begin{abstract}
A low-energy electron diffraction (LEED) apparatus which works at temperatures down to about $100 \, \mathrm{mK}$ is designed to obtain structural information of 2D helium on graphite.
This very low temperature system can be realized by reducing the thermal inflow from the LEED optics to the sample which is cooled by cryogen-free dilution refrigerator.
The atomic scattering factor of He is also estimated using a kinematical model, which suggests that  the diffraction signal from He atom can well be obtained by using a delay-line detector instead of a fluorescent screen.
\end{abstract}

\section{Introduction}
The second layer He on graphite is the ideal system to study about the 2D Fermion ($^3$He) or Boson ($^4$He) system.
By controlling the number of adsorbed atoms, one can widely change the areal density ($\rho$) and the phase from nearly ideal 2D gas to highly compressed 2D solid.
One of the interesting phase in this system is so-called the 4/7 phase, which is realized at 4/7 density ($\rho_{4/7} = 6.8 \, \mathrm{nm^{-2}}$) of the first layer\cite{Lusher1991}.
The 4/7 phase of $^3$He is considered to be an antiferromagnetic solid phase with triangle lattice and behaves as a gapless spin liquid\cite{PhysRevLett.79.3451}.
In the case of the 4/7 phase of $^4$He, 
on the other hand, an anomaly probably related to the order-disorder transition is obtained in its heat capacity at $T = 1.0 \, \mathrm{K}$\cite{VanSciver1978}.
Meanwhile, the torsional oscillator experiment shows the nonclassical rotational inertia for the 4/7 phase of $^4$He\cite{PhysRevB.53.2701}, and  theoretical calculation suggests that it is  related to the superfluidity instead of the localized solid\cite{PhysRevB.78.245414}. 
Even though the 4/7 phase of both $^3$He and $^4$He are well studied in such ways, it is still controversial what kind of state the phase is, and even the adsorbed structure had not been well understood. 
So far, only the lattice constants of monolayer commensurate and incommensurate solids were studied by neutron diffraction measurement\cite{Lauters1990}. 

To get the structural information of the 4/7 phase directly, low-energy electron diffraction (LEED) equipment works at temperatures down to $100 \, \mathrm{mK}$ (mK-LEED) is designed. 
Different from the neutron diffraction technique, one can easily distinguish the diffraction peaks of the adsorbates from those of the substrate by LEED. 
Actually, LEED at $T = 5\, \mathrm{K}$ successfully showed the detailed structure of 2D hydrogen on graphite\cite{PhysRevB.39.8628}.
To construct the mK-LEED, however, there are several issues to be considered.
First, huge thermal infrared radiation (IR) from the thermionic-emission electron gun and the LEED optics at room temperature (RT) should be minimized to obtain very low temperature.
It is also important to prevent the photodesorption of He during the measurement.
Second, the sample space with LEED should be in ultrahigh vacuum (UHV) and isolated from the vacuum for the cryogen-free dilution refrigerator (DR).
Finally, the delay-line detector (DLD) with higher resolution than a conventional  fluorescent screen is adopted because the scattering factor of He is rather small due to its large zero-point vibration. It is also helpful to minimize the power of inlet electron beam.
In this paper, we will show the instrumental design of mK-LEED and a feasibility of detection of diffraction signal from He.

\section{Instrumental design of mK-LEED}

\begin{figure}[htbp]
\begin{center}
\includegraphics[width=25pc]{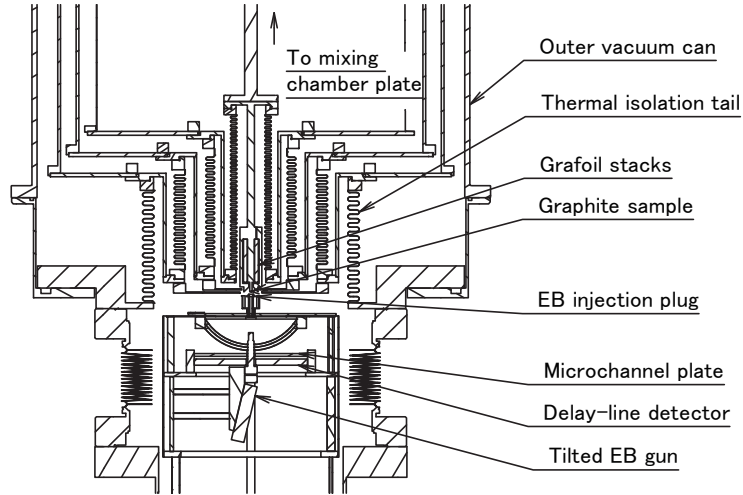}
\end{center}
\caption{The mK-LEED system which is located at the bottom of the cryogen-free DR.}
\label{overview}
\end{figure}

A cryogen-free DR \cite{DR200} with cooling power of $100 \, \mathrm{\mu W}$ at $T =100 \, \mathrm{mK}$, and the base temperature of which is $18\, \mathrm{mK}$ without any additional heat load, is adopted to cool the sample in the mK-LEED.
Since this type of DR has no outer space for cryogen and has only one vacuum space, it is easier to combine an apparatus at RT than to use the conventional DR.
The sample cell is mounted at the bottom of the mixing chamber of the DR, and the LEED optics is set at the bottom of the cryostat (Fig. \ref{overview}).

To minimize the IR from the filament of electron beam (EB) (about $150 \, \mathrm{\mu W}$), the EB gun is tilted not to face the filament directly to the sample, which can reduce the heat flow down to about $4 \, \mathrm{\mu W}$.
In addition, the LEED optics is cooled down to about $80 \, \mathrm{K}$ by liquid nitrogen, which can  reduce the IR from the optics (about $430 \, \mathrm{\mu W}$ at RT) to about $2 \, \mathrm{\mu W}$.

The UHV for LEED and the sample space is isolated from the vacuum for the DR with thin-walled ($t = 0.15 \, \mathrm{mm}$) stainless-steel bellows (the thermal isolation tail, in Fig. \ref{overview}).
This isolation system between mK regime and RT and between two types of vacuum have already been demonstrated to work very well in other experimental apparatus\cite{Kambara2007}. 

\begin{figure}[h]
	\begin{center}
\begin{minipage}{18pc}
\includegraphics[width=18pc]{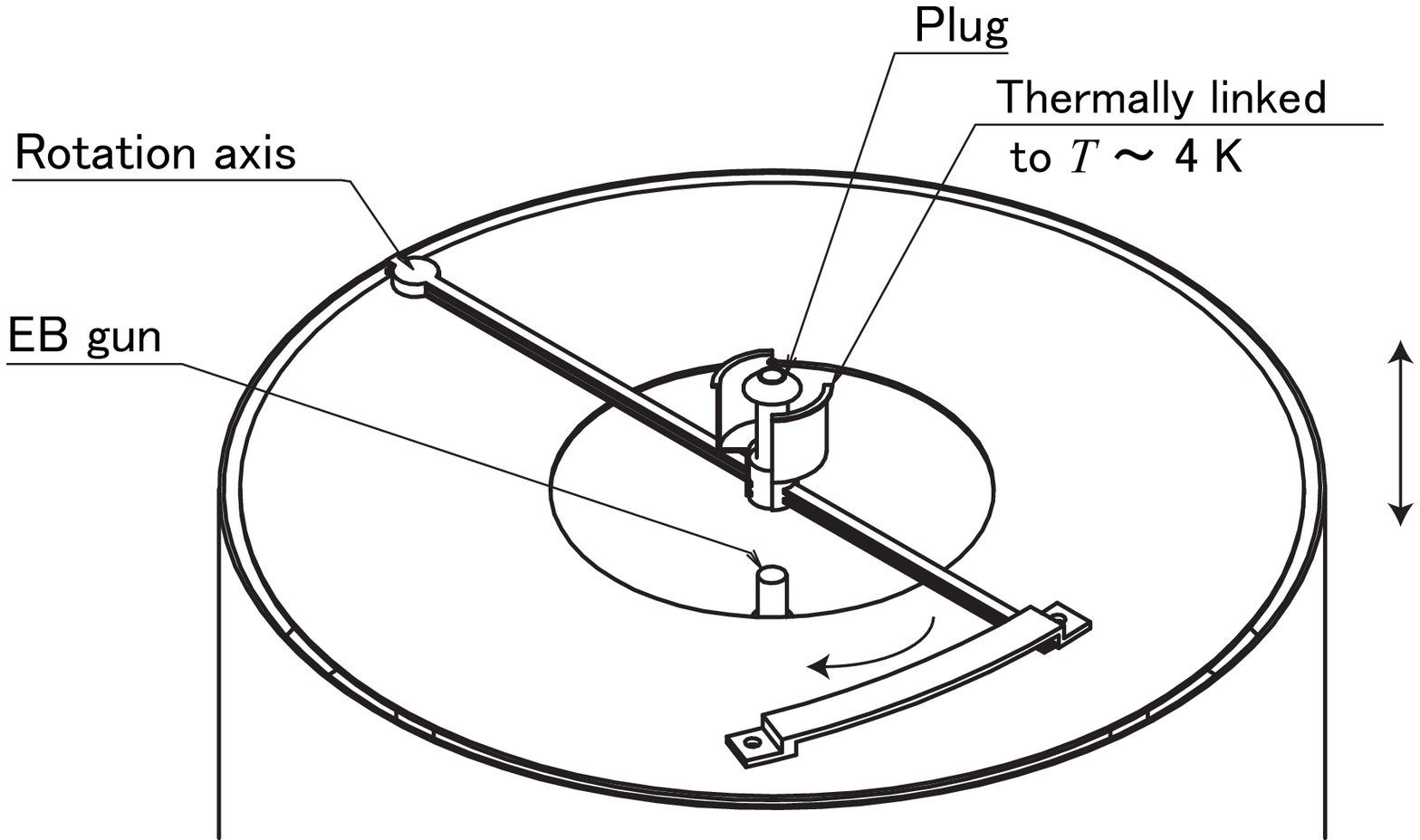}
\caption{\label{plug}The mechanism of which can linearly and rotationally movable to open/close the sample cell and electron beam.}
\end{minipage}\hspace{2pc}
\begin{minipage}{14pc}
\includegraphics[width=14pc]{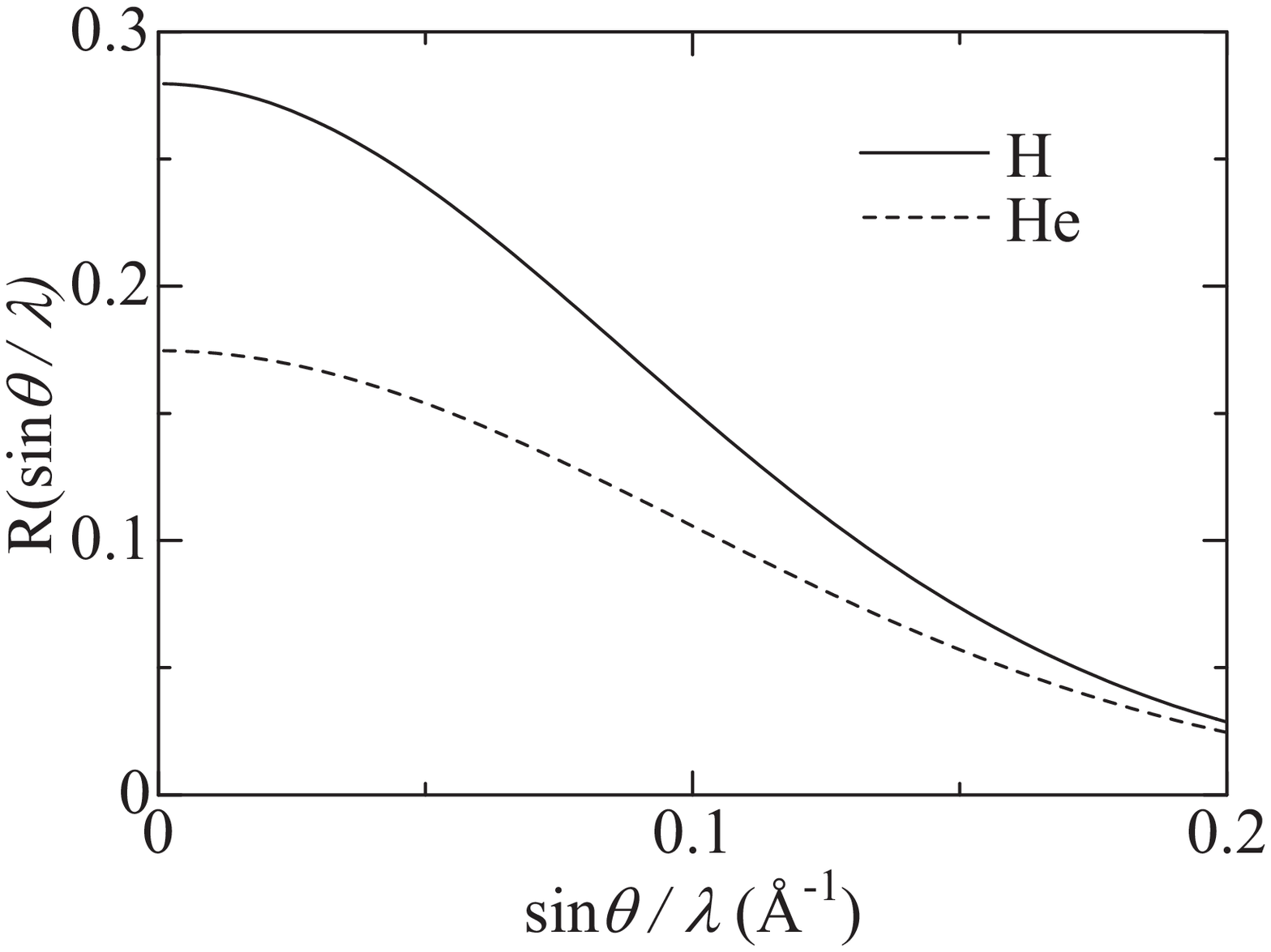}
\caption{\label{SF}Reflectance of H and He as a function of the momentum transfer.}
\end{minipage}
	\end{center}
\end{figure}

To obtain larger microcrystalline area for LEED, single-crystalline graphite with surface area of  several $\mathrm{mm^2}$ is used for the adsorption substrate.
For the precise control of the adsorption coverage, on the other hand, a stack of grafoil, an exfoliated graphite, with surface area of about $100 \, \mathrm{m^2}$ is mounted beside the adsorption substrate.
Here, one should isolate the sample cell with adsorption substrate and grafoil stack from the UHV space during the adsorbed sample preparation, while the substrate should be exposed to the LEED in UHV during the measurement. 
To realize this experimental request, a plug mechanism is designed as shown in Fig. \ref{plug}.
This plug is mounted on the shutter for EB and can move linearly and rotationally.
Therefore, it can open and close the sample cell to the UHV space.
This plug is made from a plastic with bad thermal conductivity and is thermally linked  to the $4 \, \mathrm{K}$ radiation shield at $T \approx 4 \, \mathrm{K}$  of DR when it is cooled.

Finally, estimated heat loads to the sample from each instrumental component are shown in Table 1.
The sample temperature is expected to be about $80 \, \mathrm{mK}$ and $30 \, \mathrm{mK}$ during sample preparation and the LEED measurement, respectively. 

\begin{table}[htbp]
\caption{The heat flows during sample preparation and LEED measurement. }
\label{ab}
	\begin{center}
	\begin{tabular}{ccc}
\br
&during sample preparation&LEED measurement\\
\dr
thermal isolation tail&$0.26 \, \mathrm{\mu W}$&$0.26 \, \mathrm{\mu W}$\\
electron injection plug&$76 \, \mathrm{\mu W}$&-\\
\mr
EB gun&-&$3.6 \, \mathrm{\mu W}$\\
LEED optics&-&$1.4 \, \mathrm{\mu W}$\\
\dr
total heat load&$76 \, \mathrm{\mu W}$&$5.3 \, \mathrm{\mu W}$\\
\mr
sample temperature&$80 \, \mathrm{mK}$&$30 \, \mathrm{mK}$\\
\br
\end{tabular}
	\end{center}
\end{table}

\section{Feasibility of the detection of He signal}
The vibration of the crystalline atoms reduces the scattering factor of the atoms and the amplitude of scattered electrons.
Its attenuation ratio is given by a Debye-Waller factor $\exp(-2M)=\exp(-\langle u^2\rangle \Delta k^2)$.
Here, $\langle u^2\rangle$ is mean square displacement of atom, and $\Delta k$ is the momentum transfer of the diffraction, which can be described as $\Delta k =  2 \pi \sin \theta / \lambda$ in Bragg (or Laue) condition.
Generally, atomic scattering factor against electron can be fitted by sum of five Gaussian functions with ten fitting parameters of $a_i$ and $b_i$ ($i = 1\sim 5$) \cite{Peng:zh0006}, and can be described as
\[
f(\sin \theta / \lambda )=\sum_{i=1}^5 a_i e^{-b_i (\sin \theta / \lambda)^2}.
\]
The reflectance $R(\sin \theta / \lambda)$ from a single atom can be defined by a product of the atomic scattering factor and the Debye-Waller factor
\[
R(\sin \theta / \lambda ) = \mid f(\sin \theta / \lambda )\mid ^2 e^{-2M} .
\]

In the case of He, large zero-point vibration should be considered.
The ratio of zero-point vibration amplitude to the lattice constant in 3D solid He is about 0.3\cite{H.R.Glyde1976}.
In the following calculation, we suppose this ratio of 2D He is the same as that of 3D solid He, and  $\langle u^2\rangle =1.4 \, \mathrm{\AA^2}$ for 4/7 phase.
To estimate the reflectance and check the feasibility to detect the LEED signal from He atom, we compare the $R(\sin \theta / \lambda)$ of He and H.
The $\langle u^2\rangle$ of H is estimated to be $\langle u^2\rangle =1.1 \, \mathrm{\AA^2}$ from the quantum parameter of H relative to that of He.
As a consequence, the reflectance of H and He can be estimated as shown in Fig. \ref{SF}.
The $R(\sin \theta / \lambda)$ of He atom is about $60 \, \mathrm{\%}$ smaller than that of H.

To compensate this attenuation and obtain better S/N ratio, we use a DLD for the detector of diffracted electron instead of a conventional fluorescent screen.
It is a time and position sensitive detector which converts the scattered electron to the current signal.
The sensitivity is about $10^4$ times better than that of fluorescent screen\cite{human:023302}.
Since previous studies for 2D hydrogen were succeeded with a fluorescent screen, the signal of He atoms is expected to be well detectable with DLD.
The DLD is also helpful to avoid the photodesorption of adsorbed He.
The photdesorption rate of monolayer He is reported about $10^{-3} \, \mathrm{ML/s}$ at RT and about $10^{-2}$ ML/s at $T=90 \, \mathrm{K}$\cite{Niedermayer2005}.
Since DLD is sensitive enough for LEED measurement in several seconds with EB of a few fA \cite{human:023302}, the photodesorption of He would be negligibly-small during the measurement.

In summary, we showed the design of the mK-LEED which works at temperatures down to about 100 mK.
It will, at least, be able to work at temperature lower than 300 mK, at which temperature the 4/7 phase will be realized.
Though the scattered signal will be attenuated by the large zero-point vibration of He, it could be well detectable by using the DLD. 
\section*{References}
\bibliography{ref}

\providecommand{\newblock}{}
\begin{thebibliography}{10}
\expandafter\ifx\csname url\endcsname\relax
  \def\url#1{{\tt #1}}\fi
\expandafter\ifx\csname urlprefix\endcsname\relax\def\urlprefix{URL }\fi
\providecommand{\eprint}[2][]{\url{#2}}

\bibitem{Lusher1991}
Lusher C~P, Saunders J and Cowan B~P 1991 {\em Europhysics Letters (EPL)\/}
  {\bf 14} 809

\bibitem{PhysRevLett.79.3451}
Ishida K, Morishita M, Yawata K and Fukuyama H 1997 {\em Phys. Rev. Lett.\/}
  {\bf 79} 3451

\bibitem{VanSciver1978}
{Van Sciver} S and Vilches O 1978 {\em Physical Review B\/} {\bf 18} 285

\bibitem{PhysRevB.53.2701}
Crowell P~A and Reppy J~D 1996 {\em Phys. Rev. B\/} {\bf 53} 2701

\bibitem{PhysRevB.78.245414}
Corboz P, Boninsegni M, Pollet L and Troyer M 2008 {\em Phys. Rev. B\/} {\bf
  78} 245414

\bibitem{Lauters1990}
Lauter H, Godfrin H, Frank V and Schildberg H 1990 {\em Physica B: Condensed
  Matter\/} {\bf 165} 597

\bibitem{PhysRevB.39.8628}
Cui J and Fain S~C 1989 {\em Phys. Rev. B\/} {\bf 39} 8628

\bibitem{DR200}
DR200 Cryofree \textsuperscript{\textregistered} Dilution
  Refrigerator(Oxfordshire: Oxford Instruments plc.).

\bibitem{Kambara2007}
Kambara H, Matsui T, Niimi Y and Fukuyama H 2007 {\em The Review of scientific
  instruments\/} {\bf 78} 073703

\bibitem{Peng:zh0006}
Peng L~M, Ren G, Dudarev S~L and Whelan M~J 1996 {\em Acta Crystallographica
  Section A\/} {\bf 52} 257

\bibitem{H.R.Glyde1976}
Glyde H~R 1977 {\em Rare Gas Solids\/} vol~1 (New York: Academic Press) chap~7,
  p 121

\bibitem{human:023302}
Human D, Hu X~F, Hirschmugl C~J, Ociepa J, Hall G, Jagutzki O and
  Ullmann-Pfleger K 2006 {\em Review of Scientific Instruments\/} {\bf 77}
  23302

\bibitem{Niedermayer2005}
Niedermayer T, Schlichting H, Menzel D, Payne S~H and Kreuzer H~J 2005 {\em
  Phys. Rev. B\/} {\bf 71} 045427

\end{thebibliography}
\end{document}